# Magnetic interactions and origin of high Curie temperatures in high Mn content $(Sb_2Te_3)_{1-x}(MnSb_2Te_4)_x$ quantum materials

**Candice R. Forrester,**[1,2] Christophe Testelin,[3] Kaushini Wickramasinghe,[4] David Hrabovsky,[5] Lia Krusin-Elbaum,[6,7] and Maria C. Tamargo[1,2,7]

*[1]Department of Chemistry, The City College of New York, NY, NY 10031, USA*

*[2]Ph.D Program in Chemistry, CUNY Graduate Center, NY, NY 10016, USA*

*[3]Institut des NanoScience de Paris, Sorbonne Université, CNRS, F-75005 Paris, France*

*[4]Department of Physics, University of Peradeniya, Peradeniya, 20400, Sri Lanka*

*[5]Sorbonne Université, MPBT Platform, 4 Place Jussieu, F-75005 Paris, France*

*[6]Department of Physics, The City College of New York, New York, New York 10031, USA*

*[7]PhD Program in Physics, CUNY Graduate Center, New York, New York 10016, USA*

**Abstract**

Understanding the magnetic interactions that promote high $T_C$ in topological quantum materials is essential to effectively design materials whose magnetic configuration persists at high temperatures and are thus practical to integrate into commercial spintronic devices. Here we provide evidence of the origin of the high $T_C$ observed in mixed $Mn_{1+y}Sb_{2-y}Te_4$ septuple layers and $Sb_{2-y}Mn_yTe_3$ quintuple layer structures. Analysis of the quintuple layer/septuple layer structures explored their magnetic behavior through different models of Mn-incorporation in the crystal, evidenced as Mn spin S = 5/2, and provided an explanation for the low magnetization per Mn atom observed, signature of competing ferromagnetic and antiferromagnetic interactions between $Mn^{2+}$ ions. Our studies provide insight to better understand and control the Mn incorporation in our samples to optimize their properties.

## I. Introduction

Recently our group reported the growth of $(Sb_2Te_3)_{1-x}(MnSb_2Te_4)_x$ mixed quintuple layer (QL)/septuple layer (SL) structures with Curie temperature ($T_C$) as high as 100K, the highest $T_C$ reported to date for these materials.[1] This was achieved by adjusting growth parameters: specifically, using of high Mn beam equivalent pressure (BEP) ratio and slow growth rates (GR) during the molecular beam epitaxy (MBE) growth of self-assembled mixed structures of $MnSb_2Te_4$ SLs and $Sb_2Te_3$-based QLs.[1] Adjusting these parameters resulted in increased Mn content in the crystal, particularly for mixed $(MnSb_2Te_4)_x(Sb_2Te_3)_{1-x}$ structures with $x \geq 0.7$ (>70% SL). However, evidence for how the distribution of the excess Mn atoms in the crystal leads to a higher $T_C$, or studies probing their possible magnetic interactions, particularly for samples with significantly large Mn content and very high $T_C$s, have not been provided.

In our previous work, we proposed a possible explanation for how the distribution of the excess Mn in the crystal could affect magnetic behavior.[1,2] Close analysis of derivative curves of temperature-dependent Hall effect measurements revealed the presence of two $T_C$ components with apparently distinct origins, which we suggested were the following: $T_{C1}$ is due to excess Mn in $MnSb_2Te_4$ SLs,[1,3] and $T_{C2}$ is due to high Mn content in $Sb_{2-y}Mn_yTe_3$ QL alloys.[1,2] Such alloys, although theoretically proposed, have never been experimentally demonstrated before.[4] Thus, we assert that our crystals comprise a new material: mixed $(Mn_{1+y}Sb_{2-y}Te_4)$ SLs and $(Sb_{2-y}Mn_yTe_3)$ QLs alloy structures with significant excess Mn in both the QLs and the SLs. We propose that the formation of high Mn content $Sb_{2-y}Mn_yTe_3$ QL alloys is what may lead to high $T_C$. The $Sb_{2-y}Mn_yTe_3$ QLs, although never experimentally realized, have been predicted theoretically to have such high $T_C$ values.[4] If our proposal is supported, our results represent the first experimental demonstration of this elusive metastable ternary magnetic QL alloy, whose potential high $T_C$ values have been predicted theoretically[4].

In this paper, we report detailed analyses that support our predictions. Based on energy dispersive X-ray spectroscopy (EDS) compositional data, we calculated the values of $y$ in $Sb_{2-y}Mn_yTe_3$ under different assumptions (or scenarios) as to how the excess Mn may incorporate into the crystal. We then plot the $T_C$ values obtained versus the calculated composition $y$ for each scenario and compare these plots to the theoretically predicted curves for $Sb_{2-y}Mn_yTe_3$. This comparison allows us to identify the most likely sites for Mn distribution throughout the crystal. We have also used the EDS data to calculate the magnetization per Mn atom. The data reveals very low magnetization per Mn atom, even for significant magnetic fields, suggesting opposite spin alignments and magnetic-magnetic interactions for different types of Mn substitution.

These structural and magnetic analyses lead us to conclude: 1. that there is significant Mn incorporation in the QL in the form of Mn/Sb antisites (Mn substitution in Sb sites); 2. that in the very high Mn content layers there is likely some Mn substitution for Te atoms; and 3. that the different substitutional sites for the Mn atoms present in the $Mn(Sb_{2-y}Mn_y)Te_4$ or $(Mn_{1+y}Sb_{2-y}Te_4)$ SLs (i.e., Mn in the central plane of the SL and substitutional Mn atoms) produce *ferrimagnetic* coupling of the spins within the SLs, and result in a weak average magnetization per Mn atom in the structure. Our studies provide insight to better understand and control the Mn incorporation in our samples to optimize their properties. They also help clarify the nature of the magnetic behavior and interactions present in these unique, technologically relevant materials.

**II. Experimental Details**

Thin film mixed SL/QL structures of the type $(MnSb_2Te_4)_x(Sb_2Te_3)_{1-x}$ were grown on epi-ready c-plane sapphire substrates using a Riber 2300P MBE system under ultra-high vacuum conditions ($3-5 \times 10^{-10}$ Torr). The system is equipped with reflection high-energy electron diffraction (RHEED) to monitor the growth of these materials in situ. The sapphire substrates were first thermally cleaned under vacuum to 670°C for 1 hour to remove any impurities on the surface. Fluxes for high-quality 6N antimony (Sb) with a Riber double zone cell and single zone Knudsen cells for high-quality 6N tellurium (Te) and 5N8 manganese (Mn) were measured by the BEP obtained by an ion gauge placed in the substrate position. More details of the growth can be found in refs.1 and 3.

Magneto-transport measurements were performed in the van der Pauw geometry with indium contacts using a 14 T Quantum Design physical property measurement system (PPMS) in a 1 mTorr (at low temperature) of He gas or in a Lakeshore 7600 electromagnetic system. Magnetization measurements of the samples were made using a superconducting quantum interference device (SQUID, MPMS3 from Quantum Design). X-ray EDS measurements were made in a Zeiss Supra 40 with a compact 30 mm Bruker detector.

## Results and Discussion

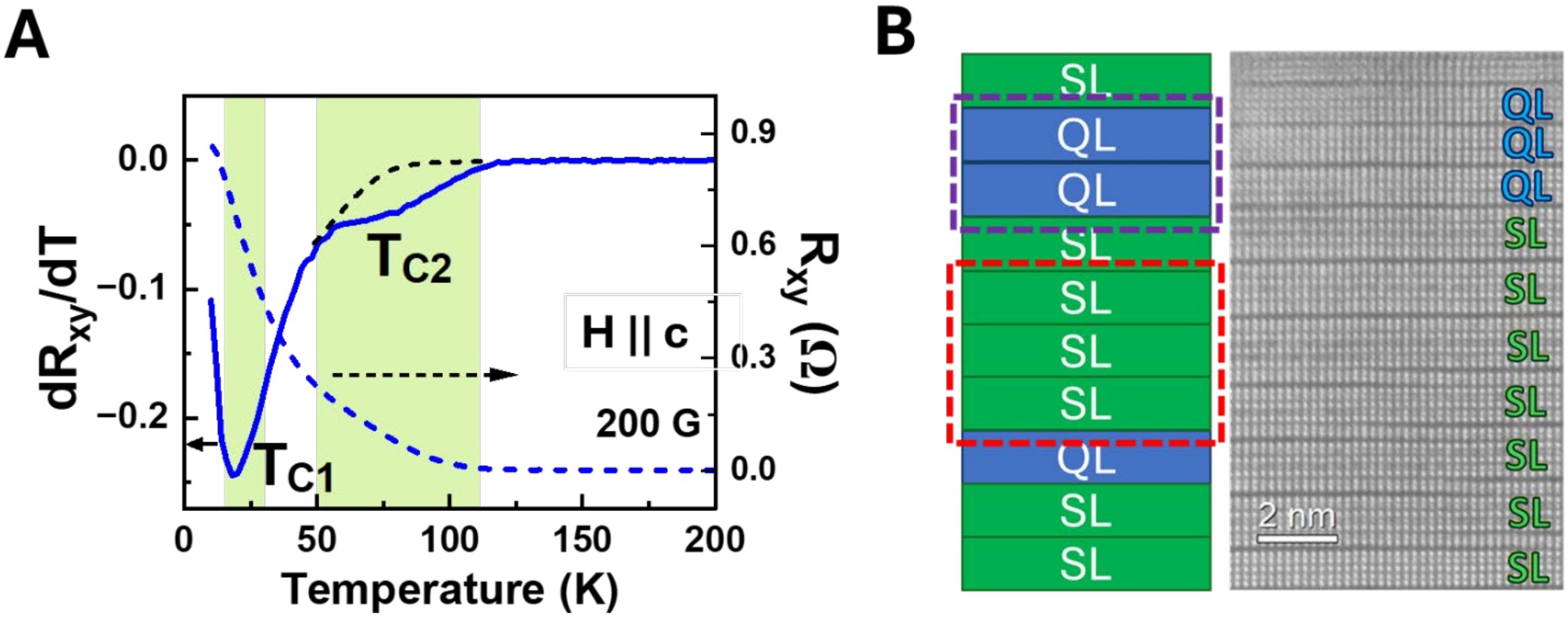

**Figure 1:** a) Temperature dependent Hall resistance (dashed) and derivative Hall resistance (solid) curves measured at 200G. Two $T_C$ values were observed. Taken from *APL Mater.* 12, 071109 (2024) b) Structural schematic of $(MnSb_2Te_4)_x(Sb_2Te_3)_{1-x}$ layers and corresponding STEM-HAADF for representative samples that were grown. Red dashed square encompasses stacked SL and the purple dashed square encompasses the Mn-rich alloy QLs.

To better understand and enhance the Curie temperature ($T_C$) of the materials, it is of interest to characterize the magnetic behavior of these complex magnetic $(MnSb_2Te_4)_x(Sb_2Te_3)_{1-x}$ topological materials. Our previous work, which conducted magnetization measurements at low fields, probed the $T_C$ in low Mn content materials, and observed good agreement for fast GR samples between Arrott plots and $R_{xy}$ plots.[2] There, a high $T_C$ of about 75K was observed. Higher $T_C$s, as high as 110K were obtained when higher Mn content was achieved in the layers.[1] We proposed that the excess Mn results in a new structure with Mn containing QL alloys: $Sb_{2-y}Mn_yTe_3$, which lead to the higher $T_C$s.[1,2]

As previously reported (seen in Figure 1a), field dependent Hall and derivative curves confirm ferromagnetic behavior with two $T_C$ components. Under an applied field parallel to the *c* axis, the Hall resistance abruptly increases below the $T_C$, supporting ferromagnetic behavior.

Derivative Hall resistance curves support the existence of two changes in slope, supporting the presence of two $T_C$s. The left panel in Figure 1b illustrates schematically the self-assembled $(MnSb_2Te_4)_x(Sb_2Te_3)_{1-x}$ thin film structures with $x \geq 0.7$, which yield the highest $T_C$. A TEM image of a grown sample with x = 0.75 (or 75%SL) presented in the right panel shows a similar distribution of SLs and QLs. We proposed that the two $T_C$ components originate from two distinct areas within the crystal.[1, 2] The $T_{C1}$, of about 20K, has been reported by others and is expected to originate from stacked $MnSb_2Te_4$ layers with excess Mn, highlighted by the red square in Figure 1b. A very high $T_{C2}$ of 70-110K originates from the QL alloy regions, highlighted by the purple square in Figure 1b. High $T_C$ QL alloys of $Sb_2Te_3$ with transition metals (TM) of the form of the $Sb_{2-y}TM_yTe_3$, have had been predicted for TM = Cr, V and Mn.[3–5] However, alloys with significant TM content have only been experimentally demonstrated for Cr and V. Our results described below are consistent with the presence, in our structures, of the high $T_C$ QL alloy $Sb_{2-y}Mn_yTe_3$ with high Mn content.

## A. Excess Mn distribution and high $T_C$ values

Previously reported EDS data indicated a very large excess of Mn in the structures with composition $(MnSb_2Te_4)_x(Sb_2Te_3)_{1-x}$ when $x \geq 0.7$, which is the compositional range in which the highest $T_C$ values are observed.[1] To understand this, we investigated plausible scenarios for the distribution of the excess Mn in these layers.

### 1. Excess Mn incorporates only into Sb sites.

We first assumed the simplest scenario, in which Mn substitutes only for Sb. It is likely that this happens both in the stoichiometric SLs ($MnSb_2Te_4$) forming $Mn_{1+y}$ $Sb_{2-y}Te_4$ and in the stoichiometric $Sb_2Te_3$ QLs forming $Sb_{2-y}Mn_yTe_3$. For simplicity, we assume that the fraction of Mn/Sb substitution (y) is the same for both the SLs and the QLs. Our EDS data presented in Ref. 1 provided values of the Mn fraction ($x_{Mn}$) and Sb fraction ($x_{Sb}$) in various samples grown with different Mn BEP ratios during growth. From the materials composition $(Mn_{1+y}Sb_{2-y}Te_4)_x$ $(Sb_{2-y}Mn_yTe_3)_{1-x}$, we can deduce the relationship between y and $x_{Mn}$ to be (see supplementary materials Section S.1.1):

$$y = [(5 + 2x)x_{Mn} - x] \qquad \text{Equation 1}$$

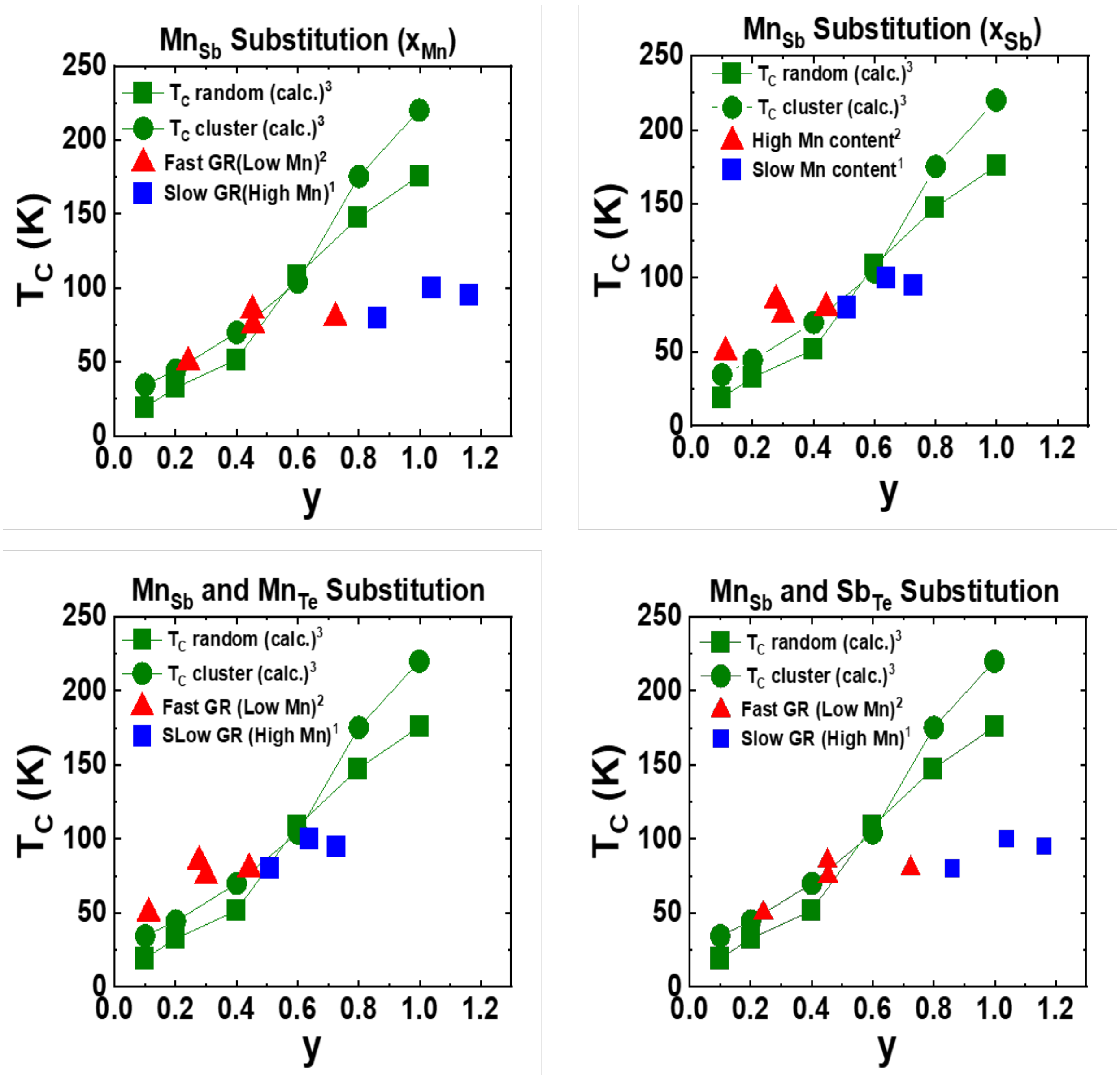

**Figure 2:** Proposed Mn distribution scenarios to best describe Mn fraction in $Sb_{2-y}Mn_yTe_3$ QL alloys. Curie temperature as a function of Mn/Sb substitution (y) for different substitution scenarios: a) Only $Mn_{Sb}$ substitution, where y is defined using $x_{Mn}$ b) Only $Mn_{Sb}$ substitution, where y is defined using $x_{Sb}$ c) $Mn_{Sb}$ and $Mn_{Te}$ substitution d) $Mn_{Sb}$ and $Sb_{Te}$ substitution.

A relationship between y and $x_{Sb}$ can also be defined:

$$y = 2 - (5 + 2x)x_{Sb} \qquad \text{Equation 2}$$

With one single type of substitution, one can see that, from the composition $(Mn_{1+y}Sb_{2-y}Te_4)_x(Sb_{2-y}Mn_yTe_3)_{1-x}$, $y$ is not unambiguous and can be obtained using the values of the SL ratio x and the Mn or Sb fractions ($x_{Mn}$ or $x_{Sb}$). The values of $T_C$ vs y for our samples were plotted using each of these equations (Figure 2a and 2b) and compared to the theoretically predicted curves.[4] Two important observations could be made: (1) different values of $y$ were obtained when $x_{Mn}$ or $x_{Sb}$ were used, and (2) the results obtained with $x_{Sb}$ are much more closely aligned with the theoretical predictions, especially for the samples with the higher $T_C$s (the slow growth rate samples). These observations suggest that our premise in Scenario 1 (one single substitution type, with only $Mn_{/Sb}$ antisites) must be incorrect, and all the Mn atoms are not substituted into Sb sites. Thus, other types of substitutional disorder must be taking place during growth. From this first scenario alone, the nature of the additional Mn induced disorder remains to be determined.

**2. Excess Mn incorporation into Sb sites and Te sites**.

Next, another occupational scenario taking place in the crystal was investigated, illustrated in Figure 2c. Careful observation of the EDS data of Ref. 1 shows that the Te content of the layer deviates from the stoichiometric quantity for samples with x > 0.7, consistent with less than stoichiometric quantities of Te in our samples. Thus, we explored the possibility that Mn is also incorporating into Te sites. This leads to a new expression for the chemical equation of our material: $(Sb_{2-y}Mn_{y+0.75z}Te_{3-0.75z})_{1-x}$ $(Mn_{1+y+z}Sb_{2-y}Te_{4-z})_x$. As seen in equation 3, z vs $x_{Te}$ is given by (see supplementary materials Section S.1.2) :

$$z = 4 - \frac{4(5+2x)}{3+x} x_{Te} \quad \text{Equation 3}$$

while y vs $x_{Sb}$ still follows eq. 2. This yields the $T_C$ vs y dependence shown on Fig. 2c with a good agreement with theory, suggesting that some Mn may be incorporating into the Te sites. Note that significant $Mn_{Te}$ antisite concentration ( ~10%) has recently been observed in MBE grown MnTe.[5]

**3. Excess Mn incorporation into Sb sites, and some Sb incorporation into Te sites.**

Another possible scenario that may explain the reduced Te content is the possibility that during growth with high Mn fluxes, as the Mn displaces Sb, the Sb may in turn be displacing Te atoms. This scenario is illustrated in Figure 2d. It should be noted that $Sb_{Te}$ antisites are a common defect for $Sb_2Te_3$ in near stoichiometric structures. These $Sb_{Te}$ antisites lead to the p-type conductivity of these layers.[6]

The expression for the chemical equation of our structures under this scenario becomes: $(Sb_{2-y+0.75z}Mn_yTe_{3-0.75z})_{1-x}(Mn_{1+y}Sb_{2-y+z}Te_{4-z})_x$. In that scenario (see supplementary materials Section S.1.3), z still is expressed using eq. 3, while y follows eq. 1. Here, the $T_C$ vs y dependence is given by the figure 2d, (identical to Fig. 2a) with a poor fit for the $T_C$ data of our samples with high $T_C$. This suggests that including Sb substitution for Te does not accurately describe our high $T_C$ values and thus is not likely occurring.

**4. Excess Mn interstitial incorporation in the VdW gap.**

Recent studies on $MnSb_2Te_4$-based crystals with high Mn contents, have considered the influence of Mn interstitials (Mn in the van der Waals gap) on the $T_C$. Those authors report $T_C \approx 70\ K$ for a Mn occupation rate $i\ \approx 30\%$ in the VdW gap.[7] One can estimate, in such scenario, what should be the Mn interstitial rate. If we modify the chemical equation of our materials to account for Mn interstitials, the equation expression becomes $Mn_i(Mn_{1+y}\ Sb_{2-y}Te_4)_x$ $(Sb_{2-y}Mn_yTe_3)_{1-x}$ with (see supplementary materials Section S.1.4):

$$i = \frac{3+x}{x_{Te}} - (5 + 2x) \qquad \text{Equation 4}$$

$$y = 2 - x_{Sb}(5 + 2x + i) \qquad \text{Equation.5}$$

The plot of $T_C$ vs $y$, using expression of y in Eq. 5 and shown in Figure 3a, presents a reasonable fit of our data with the theoretical prediction. However, we will show that this scenario can be excluded for our samples. As seen in Figure 3b, the values of $i$ for our materials were estimated, and typical slow GR samples have values of $x \approx 80\%$ and $x_{Te} \approx 51\%$ , which lead to $i \approx 80\%$ a very high interstitial rate. Such high levels of Mn in the vdW gaps should be easily observed in HR-TEM images. Numerous high-resolution HR-TEM images have been performed on our samples (see, for example, figures 1b and 3c) and no evidence of Mn interstitials in the vdW gap has ever been observed. Thus, if some Mn interstitials are present, they are in a very low concentration, and the scenario can be excluded based on the absence of interstitials in the TEM data.

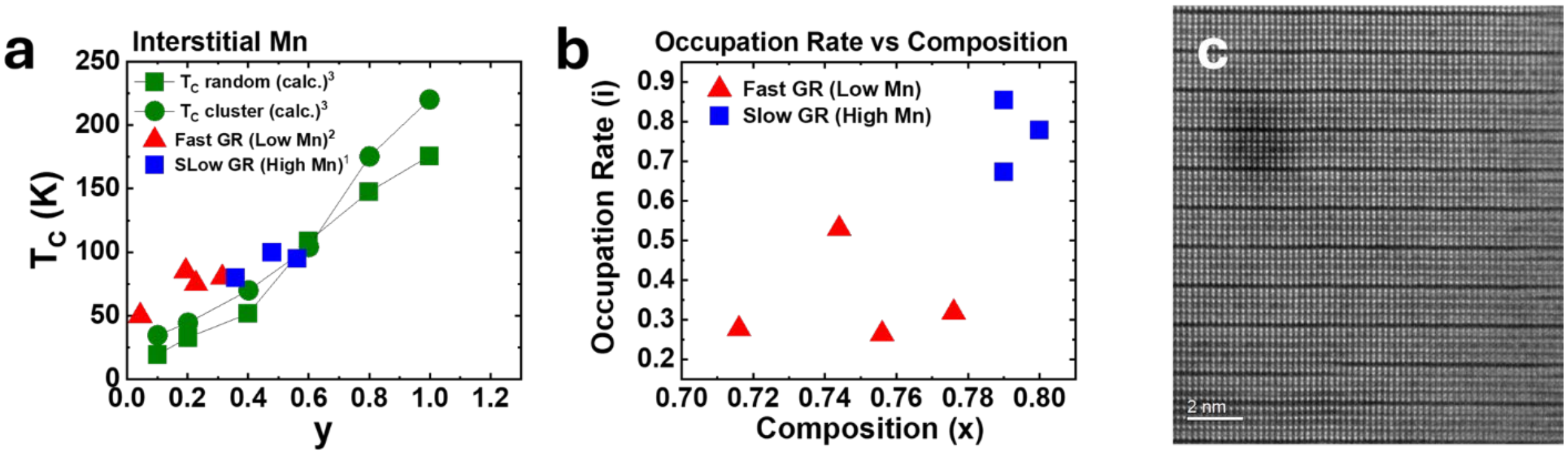

**Figure 3:** a) Proposed Mn distribution scenarios to describe Mn fraction in $(Sb,Mn)_2Te_3$ QL alloys assuming Interstitial Mn occupation b) Occupancy disorder as a function of composition. Slow GR, high Mn content materials are expected to have significant rates of occupancy disorder in the interstitial regions. c) shows TEM that does not show Mn atoms in the VdW gaps.

The excellent fit for scenario 2 to our data supports our hypothesis that the high $T_C$ component in the materials originates from the ($Sb_{2-y}Mn_yTe_3$) QL layer alloys. The results further suggest that the materials have a composition given by the equation $(Sb_{2-y}\,Mn_{y+0.75z}\,Te_{3-0.75z})_{1-x}$ $(Mn_{1+y+z}Sb_{2-y}Te_{4-z})_x$. which includes both $Mn_{Sb}$ and $Mn_{Te}$ antisites. This analysis also suggests that the magnetic properties of these materials may be highly complex in nature due to the different types of magnetic ions and magnetic interactions likely present in the material. This will be addressed in the following sections.

**B. Analysis of the magnetization and magnetic interactions**

**1. Curie-Weiss Law: Effective Magnetic Moment**

The effective magnetic moment $\mu_{eff}$ is an important parameter in the magnetization behavior and is a function of the magnetic ion state. In our compounds, the Mn ions are expected to be in a high spin state (S = 5/2 for $Mn^{2+}$). Nonetheless, lower spin states ($Mn^{3+}$, S = 2) have also been discussed, especially in QL compounds.[8,9] To extract $\mu_{eff}$, we have measured the magnetic susceptibility $\chi(T)$ versus temperature and considered the Curie-Weiss (CW) law. This law is a fundamental way to study the intrinsic moment of magnetic ions, in this case $Mn^{2+}$, in the paramagnetic regime. By assuming each magnetic moment is subject to a mean field induced by interaction with other spins, one deduces the average magnetization per ion and its temperature dependence. Here, two kinds of samples are considered under this modeling: (i) one fast GR sample with a single $T_C$, $x = 66.4\%$ and $x_{Mn} = 11.8\%$, (ii) one slow GR sample with two $T_C$s, $x = 81\%$ and $x_{Mn} = 27.8\%$.

For the sample with a single $T_C$, the CW law can be written as:

$$\chi(T) = \chi_0 + \frac{C}{T - T_C} \qquad \text{Equation 6}$$

with $\chi_0$, a constant associated to the diamagnetic contribution from the sapphire substrate. $T_C$ is the CW temperature, and $C$ is the CW constant given by:

$$C = \frac{n_{Mn}\mu_{eff}^2}{3k_B} \qquad \text{Equation 7}$$

with $n_{Mn}$ the Mn volumetric concentration and $k_B$ the Boltzman constant. Here, $n_{Mn}$ can be estimated from the Mn composition and QL/SL ratio (see supplementary materials Section S.2.). Fig. 5a shows the temperature dependence of $\chi(T)$ for the sample with a single $T_C$, and the fitting of the CW law using eq. 6. The CW constant can also be estimated from the high temperature behavior of the product $(\chi(T) - \chi_0).T$ (see S.I. S.1). Combining the two approaches, one deduces C = (2.5 ± 0.3) $10^{-2}$ emu.$Oe^{-1}$.$cm^{-3}$, leading to a value of $\mu_{eff} = (5.55 \pm 0.35)\ \mu_B$.

For the second sample, with two $T_C$ contributions, the magnetic susceptibility must be written:

$$\chi(T) = \chi_0 + \frac{C_1}{T - T_{C_1}} + \frac{C_2}{T - T_{C_2}} \qquad \text{Equation 8}$$

Fig. 5b presents the magnetic susceptibility $\chi(T)$ versus temperature, compared to the fitting using Eq. 8. From this curve, one gets the full contribution $C = C_1 + C_2 = 5.5 \pm 0.4\ 10^{-2}$ emu.Oe$^{-1}$.cm$^{-3}$. From the high temperature behavior of $(\chi(T) - \chi_0).T$, one estimates an upper bound for $C = 6.5 \pm 0.5\ 10^{-2}$ emu.Oe$^{-1}$.cm$^{-3}$K$^{-1}$ (see Section S.2 in supplementary materials). One then deduces C = (6.0 ± 0.5) $10^{-2}$ emu.Oe$^{-1}$.cm$^{-3}$K$^{-1}$, and $\mu_{eff} = (5.59 \pm 0.22)\ \mu_B$. The $T_C$ values obtained from the CW law fits are very consistent with the ones deduced from the anomalous Hall resistance dependence in temperature.

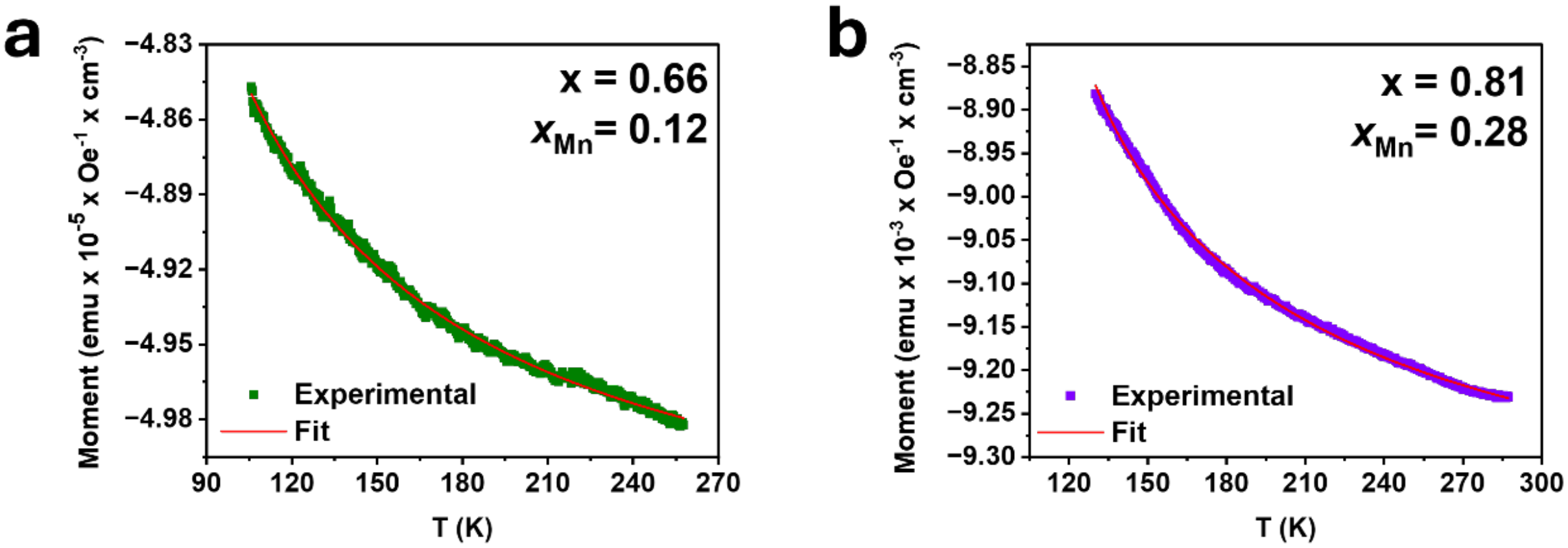

**Figure 4:** Magnetic susceptibility χ(T) versus temperature for (a) sample with a single $T_C$ temperature: $T_C$ = 19.9 ±0.5 K and C = 2.20 ± 0.2) $10^{-2}$ emu.Oe$^{-1}$.cm$^{-3}$ ; (b) sample with two $T_C$ temperatures : $T_{C_1} = 19.2 \pm 1$ K and $T_{C_2} = 97 \pm 4$ K ; $C_1 = 5.1 \pm 0.3\ 10^{-2}$ and $C_2 = 3.7 \pm 0.7\ 10^{-3}$ emu.Oe$^{-1}$.cm$^{-3}$.

The effective magnetic moments, obtained from experimental data, for the two types of samples can be compared to the theoretical value, for a spin S, and given by:

$$\mu_{eff} = g\sqrt{S(S+1)}\ \mu_B \qquad \text{Equation 9}$$

where g is the Landé factor and S is the total angular momentum quantum number. In the case of $Mn^{2+}$, g = 2 and S = 5/2. Eq. 9, leads then to $\mu_{eff} = 5.92\ \mu_B$, a value close to the experimental ones, which confirms that the Mn ions are dominantly in the high spin state. We will then assume S = 5/2 spin states for the magnetization analysis in the next section.

**2. Magnetization per Mn atom: Role of spin-spin interactions**

We then considered the magnetization (M) behavior at low temperature and high field (up to 40 kOe), for a sample with $x = 0.81$ and $x_{Mn} = 0.28$ and calculated the magnetization per Mn atom, $M_{Mn}$. Using the film volume V, of 1.13x$10^{-7}$ $cm^{-3}$ (based on sample surface area and thickness) and the Mn concentration $x_{Mn}$, one first calculated the volume magnetization, $m = \frac{M}{V}$. With the average unit cell volume $\Omega_a$ of 0.199x$10^{-21}$ $cm^{-3}$ (calculated from Eq. S4 in supplementary materials) and the average number of Mn atoms in a unit cell, $\eta = x_{Mn}(5 + 2x) = 1.84 \pm 0.07$, one obtains the volume concentration of Mn atoms, $n_{Mn}$ = 9.25 ± 0.25x$10^{21}$ Mn $cm^{-3}$, (see Eq. S3). Finally, $M_{Mn}$ is determined using:

$$M_{Mn} = \frac{m}{n_{Mn}} \qquad \text{Equation 10}$$

Figure 5a shows the field cooled (FC) magnetization $M_{Mn}$ versus temperature, under a 200 Oe field. The remanent magnetization measured after turning off the magnetic field is of the same order as the FC $M_{Mn}$, with low and high temperature contributions, signature of two TCs. A clear hysteresis is observed at T = 5 K (fig. 5b), signature of ferromagnetic coupling. Figure 5c evidences the magnetization up to 40 kOe, at variable temperatures from below to above the two TCs.

Unlike the CW analysis, in the low temperature regime (below paramagnetic regime), the magnetic interactions cannot be analyzed in a mean field approach: ferromagnetic or antiferromagnetic interaction are inducing parallel or antiparallel spin alignment, and existence of remanent magnetization.

From Figure 5, one measures the average remanent or high field magnetic moment per Mn atom at T = 4K, respectively 0.19 ± 0.02 $\mu_B$ and 0.87 ± 0.08 $\mu_B$. One clearly observes high field values far from saturation $M_{Mn} = 5\mu_B$, expected for $Mn^{2+}$ spin $S = 5/2$ (as previously determined from CW law). These low values are evidence of antiferromagnetic and ferromagnetic couplings. To understand these low $M_{Mn}$ experimental values, we assumed various spin alignments in the structure depending on the position of the Mn in the lattice for the Mn incorporation scenario 2 (the only scenario consistent with our experimental data) described above (see Section 1.2 in supplementary materials).

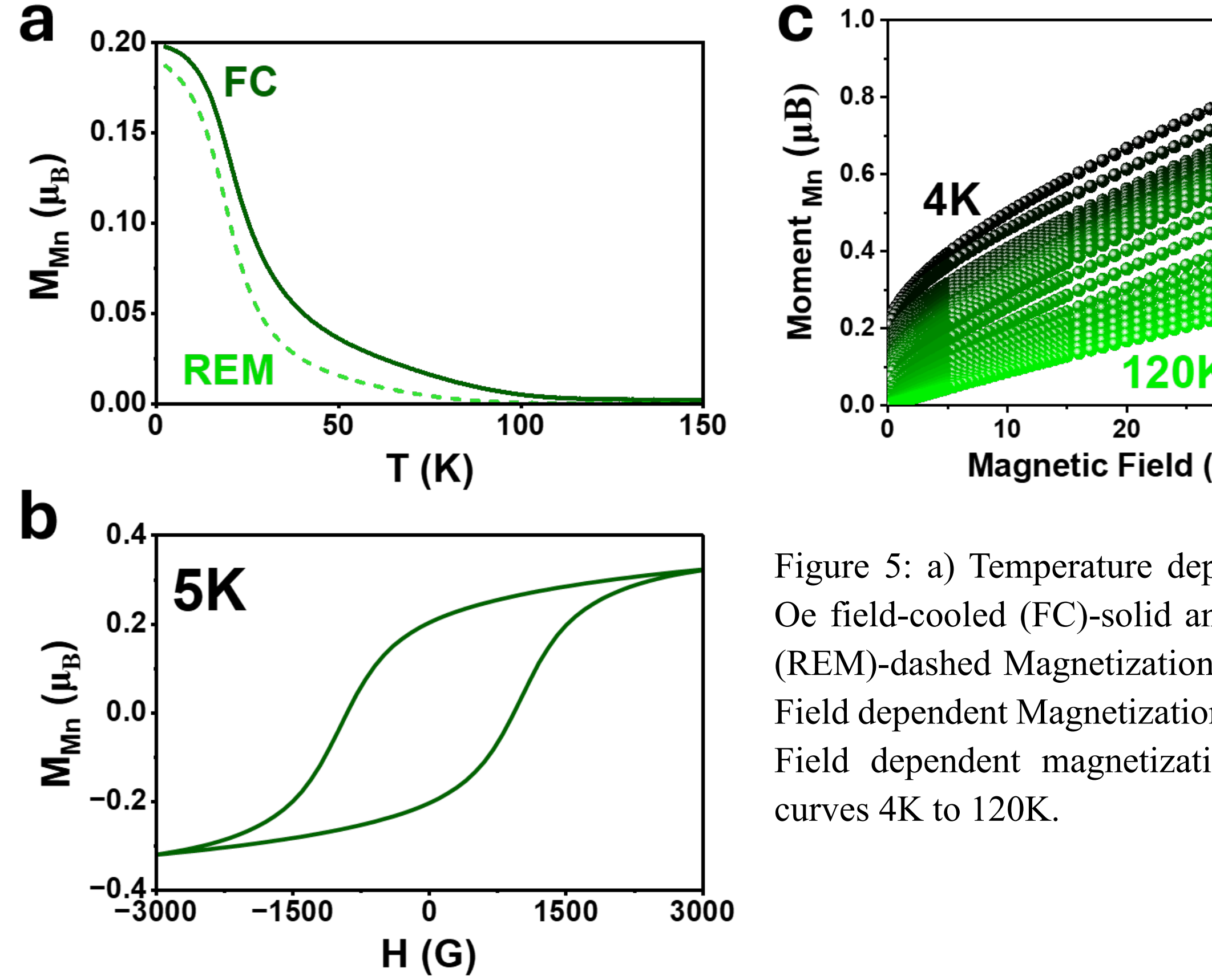

Figure 5: a) Temperature dependent 200 Oe field-cooled (FC)-solid and remanent (REM)-dashed Magnetization per Mn. b) Field dependent Magnetization per Mn. c) Field dependent magnetization $M_{Mn}(H)$ curves 4K to 120K.

In scenario 2, illustrated Figure 2c, Mn is expected to substitute into Sb and Te sites resulting in the molecular formula, $(Sb_{2-y}Mn_{y+0.75z}Te_{3-0.75z})_{1-x}$ $(Mn_{1+y+z}Sb_{2-y}Te_{4-z})_x$. Equations 11 and 12 describe the Mn incorporation in the SLs and in the QLs, respectively.

$$(1+y+z)(x) \quad \text{Equation 11}$$

$$(y+0.75z)(1-x) \quad \text{Equation 12}$$

| Spin orientation | $M_{Mn}$ (SL) | $M_{Sb}$ (SL) | $M_{Te}$ (SL) | $M_{Sb}$ (QL) | $M_{Te}$ (QL) | $M_{Mn}$ ($\mu_B$) |
|---|---|---|---|---|---|---|
| A | +1 (↑) | -1 (↓) | -1 (↓) | +1 (↑) | -1 (↓) | 0.1 ± 0.1 |
| B | +1 (↑) | -1 (↓) | +1 (↑) | +1 (↑) | +1 (↑) | 2.2 ± 0.2 |
| C | +1 (↑) | -1 (↓) | | +1 (↑) | | 1.15 ± 0.15 |

Table I: Possible spin orientations for scenario 2 of Mn incorporation in $(Sb_{2-y}Mn_{y+0.75z}Te_{3-0.75z})_{1-x}$ $(Mn_{1+y+z}Sb_{2-y}Te_{4-z})_x$ , used to calculate $M_{Mn}$, the average magnetization per Mn. For each scenario, the magnetization was calculated based on possible Mn spin orientations, either spin up (+1 /↑) or spin down (-1 /↓). Spin orientation A yields a magnetization per Mn atom ($M_{Mn}$) closest to experimental magnetization curves. Spin orientation C neglects the contribution of $Mn_{Te}$ atoms.

The y and z values are calculated using eq. 2 and 3, respectively (x = 0.81). There are three spin orientations (spin orientations A, B and C) that can be considered based on different Mn substitution sites, shown in Table I.

$M_{Mn}$ was calculated considering 5 $\mu_B$ magnetic moments (spin S = 5/2), the spin alignments (see Table I) and the proportion and weight of the Mn in the structure, based on the molecular formula expressing the way Mn is incorporated into the QLs and SLs. Different parallel and antiparallel configurations are presented in Table I, with spin-orientation coefficients: spin up (+1) and spin down (-1).

Table I summarizes the estimated average magnetization per Mn for the possible spin orientations A through C, resulting from the Mn incorporation in scenario 2. We have included a 10% error for $y$ and $z$ due to EDS uncertainty and calculated $M_{Mn}$ to be 0.1 ± 0.1 , 2.2 ± 0.2 and 1.15 ± 0.15 $\mu_B$, respectively. All magnetic moment values obtained are lower than the expected values for saturated $Mn^{2+}$ spin. Spin-orientation A has the closest value to the extracted experimental value, 0.19 ± 0.02 $\mu_B$ from $M_{rem}$ curves. Good agreement between these calculated and experimental values provides more evidence for the low average magnetic moments, indicating ferrimagnetic-like behavior. The structural disorder and corresponding magnetic disorder clearly promote some antiferromagnetic couplings between neighboring $Mn^{2+}$ ions. Mechanisms such as super-exchange explain possible origins for these couplings happening between two cations ($Mn^{2+}$) over an intermediate anion (in this case $Te^{2-}$). A closer look at the orbitals of $Mn^{2+}$ reminds us that it has a half-filled d-orbital shell, with all the spins aligned. When the d-orbitals of $Mn^{2+}$ and p-orbital of $Te^{2-}$ overlap and hybridize, virtual electron transfer happens between the anion and cation. However, to ensure that two electrons do not occupy the same quantum state, the hopping electron from the anion pairs are in an antiparallel configuration to the cation electron to avoid violating the Pauli exclusion principle.

## IV. Conclusions

In this paper we presented evidence to support our predictions of the formation of a new topological material: $(Mn_{1+y}Sb_{2-y}Te_4)_x(Sb_{2-y}Mn_yTe_3)_{1-x}$, and more specifically, the first experimental realization of the QL alloy $Sb_{2-y}Mn_yTe_3$ with high Mn content.. The high $T_C$ values observed as well as the high Mn content imply that these $Sb_{2-y}Mn_yTe_3$ QLs have Mn fractions as high as y = 0.6, much higher than ever experimentally realized. Such high y values may be possible due to the far from equilibrium MBE growth conditions. To understand how the excess

Mn is incorporated, various scenarios for the Mn distribution were explored. Our data suggests that excess Mn is incorporating into Sb sites in both the QL and SL layers; some Mn is also displacing Te atoms, consistent with the reduction of Te fraction in the high Mn containing samples seen in the EDS data. Interstitial Mn incorporation was also investigated. Despite reported arguments for interstitial Mn in these materials, our results do not support or show evidence of interstitial Mn. Our studies strongly support that the main contributor to the high $T_C$ in our structures was the $Mn_{Sb}$ substitution in $Sb_{2-y}Mn_yTe_3$ QLs, in agreement with theoretical predictions.

We also investigated the magnetic nature of this materials system. The magnetic behavior at high temperature was probed based on Curie-Weiss law and yielded effective magnetic moment values $\mu_{eff}$ in very good agreement with $Mn^{2+}$ spin S = 5/2. Significantly low values of magnetization per Mn atom was extracted from experimental magnetization curves compared to the expected saturated magnetization. To explain this low magnetization, different possible spin alignments for the different types of Mn substitutions were considered, assuming scenario 2, where Mn atoms incorporated into Sb and Te sites, best describes our system. Three possible spin orientations were proposed, with different ferromagnetic and antiferromagnetic coupling between Mn spins. Estimates of the magnetization per Mn atom for these three spin orientations showed that ferrimagnetic alignment of the spins can explain the low $M_{Mn}$ values.

Clearly, our investigations show that these materials contain a large amount of disorder, in the form of excess Mn in different lattice sites, in contrast to the case of the stoichiometric materials. This suggests short-range interactions such as super-exchange or even Kinetic exchange may be present. Published literature suggests the deviation from ideal effective magnetic moment can possibly be influenced by impurities, canting or randomly oriented domains. [10–12]

**Acknowledgements**

This work was supported by NSF Grant no. DMR-2011738 (NSF MRSEC PAQM) and NSF Grant no. HRD-2112550 (Phase II CREST IDEALS). This work was also supported by Project DYN- TOP ANR-22-CE30-0026-01 and Project IRP THERMOSPIN CNRS-CCNY-Sorbonne University. The authors would like to acknowledge the Nanofabrication Facility of the CUNY Advanced Science Research Center (ASRC) for instrument use and scientific and technical assistance and the staff of the MPBT (physical properties—low temperature) platform of Sorbonne Université for their support.

## Supplementary Material

### S.1 Excess Mn distribution and substitution rates

Different atom substitution and/or interstitial scenarios are considered, leading to different compound compositions. Each substitution rate is assumed identical in septuple and quintuple layers.

#### S1.1. Excess Mn incorporates only into Sb sites.

$(Sb_{2-y}Mn_yTe_3)_{1-x}\ (Mn_{1+y}Sb_{2-y}Te_4)_x$

| | Average number of atoms per unit cell | Atom fraction |
|---|---|---|
| All atoms | $5+2x$ | |
| Te | $3+x$ | $x_{Te}=\frac{3+x}{5+2x}$ |
| Sb | $2-y$ | $x_{Sb}=\frac{2-y}{5+2x}$ |
| Mn | $x+y$ | $x_{Mn}=\frac{x+y}{5+2x}$ |

Two different values are possible for $y$ :

$$y=(5+2x)x_{Mn}-x \qquad \text{and} \qquad y=2-(5+2x)x_{Sb}$$

#### S1.2. Excess Mn incorporation into Sb sites and Te sites.

$(Sb_{2-y}\ Mn_{y+0.75z}\ Te_{3-0.75z})_{1-x}\ (Mn_{1+y+z}Sb_{2-y}Te_{4-z})_x$

| | Average number of atoms per unit cell | Atom fraction |
|---|---|---|
| All atoms | $5+2x$ | |
| Te | $(1-\frac{z}{4})(3+x)$ | $x_{Te}=(1-\frac{z}{4})\frac{(3+x)}{5+2x}$ |
| Sb | $2-y$ | $x_{Sb}=\frac{2-y}{5+2x}$ |
| Mn | $x+y+\frac{z}{4}(3+x)$ | $x_{Mn}=\frac{x+y+\frac{z}{4}(3+x)}{5+2x}$ |

One gets $\quad y=2-(5+2x)x_{Sb} \quad$ and $\quad z=4-\frac{4(5+2x)}{3+x}x_{Te}$

### S1.3. Excess Mn incorporation into Sb sites, and some Sb incorporation into Te sites.

$(Sb_{2-y+0.75z}Mn_yTe_{3-0.75z})_{1-x}(Mn_{1+y}Sb_{2-y+z}Te_{4-z})_x$

| | Average number of atoms per unit cell | Atom fraction |
|---|---|---|
| All atoms | $5+2x$ | |
| Te | $(1-\frac{z}{4})(3+x)$ | $x_{Te}=(1-\frac{z}{4})\frac{(3+x)}{5+2x}$ |
| Sb | $2-y+\frac{z}{4}(3+x)$ | $x_{Sb}=\frac{2-y+\frac{z}{4}(3+x)}{5+2x}$ |
| Mn | $x+y$ | $x_{Mn}=\frac{x+y}{5+2x}$ |

One gets $y=(5+2x)x_{Mn}-x$ and $z=4-\frac{4(5+2x)}{3+x}x_{Te}$

### S1.4. Excess Mn interstitial incorporation in the VdW gap

$Mn_i(Mn_{1+y}Sb_{2-y}Te_4)_x(Sb_{2-y}Mn_yTe_3)_{1-x}$

| | Average number of atoms per unit cell | Atom fraction |
|---|---|---|
| All atoms | $5+2x$ | |
| Te | $3+x$ | $x_{Te}=\frac{3+x}{5+2x+i}$ |
| Sb | $2-y$ | $x_{Sb}=\frac{2-y}{5+2x+i}$ |
| Mn | $x+y+i$ | $x_{Mn}=\frac{x+y+i}{5+2x+i}$ |

One gets $y=2-x_{Sb}(5+2x+i)$ and $i=\frac{3+x}{x_{Te}}-(5+2x)$

**S.2.Effective Magnetic Moment: Curie-Weiss Law**

We calculated effective magnetic moments to extract information about the inherent magnetic interactions in our material. To determine the effective magnetic moment, the Curie-Weiss law equation was first used to extract the Curie constant, *C* within the paramagnetic regime as seen in Figure 5. By assuming each magnetic moment is subject to a mean field induced by coupling to other spins, one can deduce the average magnetization per ion and then the volumetric magnetic susceptibility, written as :

$$\chi(T) = \chi_0 + \frac{C}{T-T_C} \qquad \text{Equation S1}$$

$\chi_0$ is the substrate magnetic susceptibility. C the CW constant defined by :

$$C = \frac{n_{Mn}\mu_{eff}^2}{3k_B} \qquad \text{Equation S2}$$

with $\mu_{eff}$ is the effective magnetic moment and can be obtained by measuring C and an estimation of the Mn volume density $n_{Mn}$, given by :

$$n_{Mn} = \frac{(5+2x).x_{Mn}}{\Omega_a} \qquad \text{Equation S3}$$

where $(5 + 2x)$ is the average atom number per unit cell, $x_{Mn}$, the probability of the atom being Mn (extracted from EDS measurement) and $\Omega_a$ is the average unit cell volume defined by:

$$\Omega_a = x\Omega_{SL} + (1-x)\Omega_{QL} \qquad \text{Equation S4}$$

with $\Omega_{SL} = 0.208\, nm^{-3}$ and $\Omega_{QL} = 0.160\, nm^{-3}$ , the SL and QL unit cell volumes, respectively. For the first sample ($x = 66.4\%$ and $x_{Mn} = 11.8\%$), $\Omega_a = 0.191\, nm^{-3}$ and $n_{Mn} =$ $3.90\ 10^{21}\ cm^{-3}$. For the second sample ($x = 81\%$ and $x_{Mn} = 27.8\%$), $\Omega_a = 0.199\ nm^{-3}$ and $n_{Mn} = 9.25\ 10^{21}\ cm^{-3}$.

In order to estimate the CW constants, we have used eqs. 6 or 8 for our samples (see main text and figure 5), but we have also plotted the product $(\chi(T) - \chi_0).T$ versus T. This function is supposed to decrease and tend to C for large temperatures. As shown in Fig. S1, one observes a plateau at high temperature. For the two considered samples, this corresponds to CW constants $C = 2.80 \pm 0.3\ 10^{-2}$ and $6.5\ \pm 0.5\ 10^{-2}$ emu.Oe$^{-1}$.cm$^{-3}$K$^{-1}$, respectively.

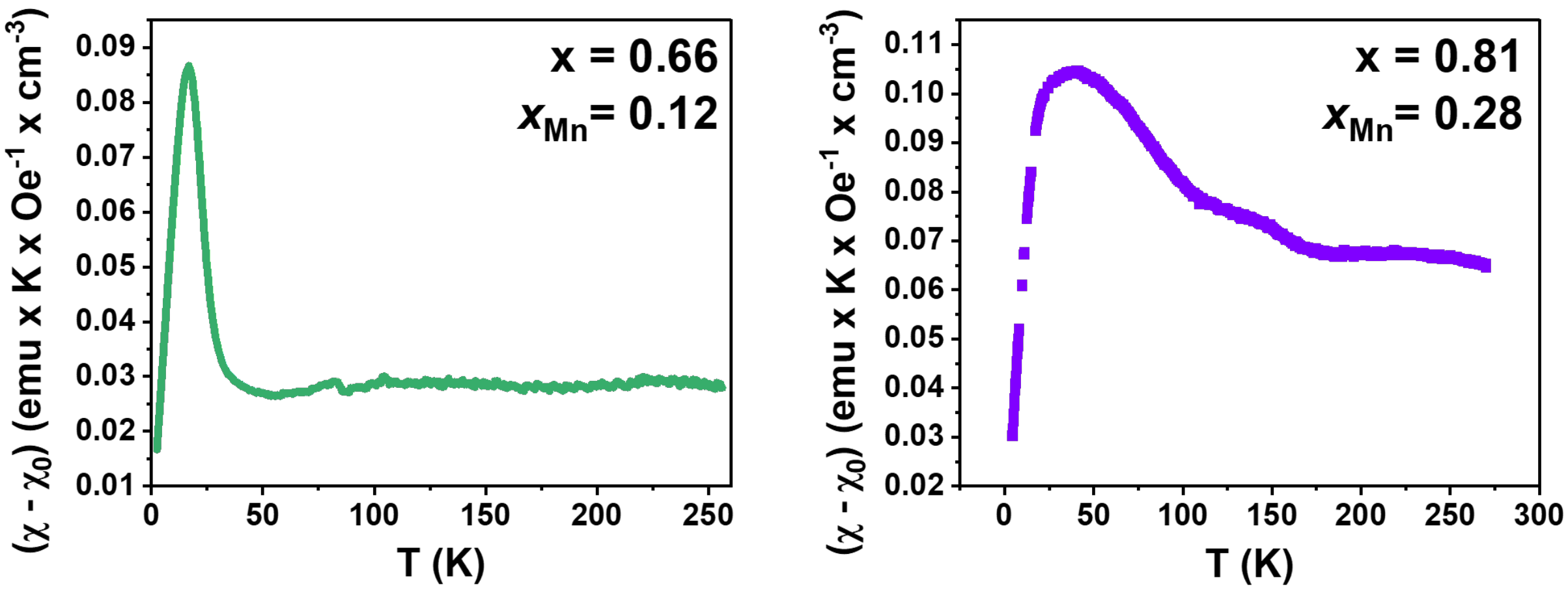

**Figure S1.** Temperature dependence of the product $(\chi(T) - \chi_0)T$ for (a) sample with a single CW temperature; (b) sample with two CW temperatures.

Associated to the CW law fitting (figure 4), they lead to the CW constant ranges $C = 2.5 \pm 0.3\ 10^{-2}$ and $= 6.0 \pm 0.5\ 10^{-2}$ emu.Oe$^{-1}$.cm$^{-3}$K$^{-1}$, respectively. The effective magnetic moment $\mu_{eff}$ can then be calculated using eq S2-4 and compared to $\mu_B$ ($\mu_B = 9.274 \times 10^{-21}$ emu), as discussed in the main text.